\documentclass[printer]{aa}
\usepackage{epsfig,times}

\def \ref {\noindent\hangindent=1.0in\hangafter=1}

\def\ltsima{$\; \buildrel < \over \sim \;$}
\def\simlt{\lower.5ex\hbox{\ltsima}} % < over ~
\def\gtsima{$\; \buildrel > \over \sim \;$}
\def\simgt{\lower.5ex\hbox{\gtsima}} % > over ~

\begin{document}

\title{Broad-band continuum and line emission of the $\gamma$-ray blazar PKS~0537--441}

\authorrunning{E. Pian et al.} 
\titlerunning{Broad-band spectrum of PKS~0537--441} 

\author{E. Pian\inst{1}, 
R. Falomo\inst{2},
R. C. Hartman\inst{3},
L. Maraschi\inst{4},
F. Tavecchio\inst{4},
M. Tornikoski\inst{5},
A. Treves\inst{6},
C. M. Urry\inst{7},
L. Ballo\inst{4}, 
R. Mukherjee\inst{8},
R. Scarpa\inst{9},
D. J. Thompson\inst{3},
J. E. Pesce\inst{10}}

\institute{Osservatorio Astronomico di Trieste, Via G.B. Tiepolo 11, I-34131
Trieste, Italy
\and
Osservatorio Astronomico di Padova, via dell'Osservatorio 5, I-35122
Padova, Italy
\and
NASA Goddard Space Flight Center, Greenbelt, MD 20771, USA
\and
Osservatorio Astronomico di Brera, via Brera 28, I-20121 Milano, Italy
\and
Metsahovi Radio Observatory, Metsahovintie 114, FIN--02540 Kylmala, Finland
\and
Dipartimento di Scienze, University of Insubria, Via Valleggio 11, I-22100 Como,
Italy
\and
Department of Physics, Yale University, P.O. Box 208121, New Haven CT 06520-8121
\and
Department of Physics and Astronomy, Barnard College, Columbia University, New
York, NY 10027
\and 
European Southern Observatory, 3107 Alonso de Cordova, Santiago, Chile
\and
Eureka Scientific and George Mason University, 4400 University Drive,
Fairfax, Virginia, VA 22030
}

\offprints{E. Pian \\ \email{pian@ts.astro.it}}   

\abstract{PKS~0537--441, a bright $\gamma$-ray emitting blazar, was observed at
radio, optical, UV and X-ray frequencies during various EGRET pointings, often
quasi-simultaneously. In 1995 the object was found in an intense emission state at
all wavelengths. BeppoSAX observations made in 1998, non-simultaneously with
exposures at other frequencies, allow us to characterize precisely the spectral
shape of the high energy blazar component, which we attribute to inverse Compton
scattering. The optical-to-$\gamma$-ray spectral energy distributions at the
different epochs show that the $\gamma$-ray luminosity dominates the bolometric
output. This, together with the presence of optical and UV line emission, 
suggests that, besides the synchrotron self-Compton mechanism, the Compton
upscattering of photons external to the jet (e.g., in the broad line region) may
have a significant role for high energy radiation.  The multiwavelength
variability can be reproduced by changes of the plasma bulk Lorentz factor. The
spectrum secured by IUE in 1995 appears to be partially absorbed shortward of
$\sim$1700 \AA.  However, this signature is not detected in the HST spectrum taken
during a lower state of the source.  The presence of intervening absorbers is not
supported by optical imaging and spectroscopy of the field. }

\maketitle 

\section{Introduction}

Radiation in the MeV-GeV energy range has been firmly detected by EGRET in 65
active galactic nuclei, all of blazar type (von Montigny et al. 1995;  Hartman
1999; Hartman et al. 1999).  The $\gamma$-ray emission, coupled to the small
emitting volumes inferred by variability time scales, implies beaming in a
relativistic jet (which makes the emitting regions transparent to $\gamma$-ray
photons, e.g., McBreen 1979; Maraschi et al.  1992), a distinctive
characteristic of blazars.  Among EGRET blazars, many exhibit variability on a
range of timescales from years to days (Wehrle et al.  1998;  Mattox et al. 1997;
Bloom et al. 1997; Hartman 1996; Hartman 1999).

There is a general consensus that the emission of blazars at energies higher than
10 MeV should be attributed to inverse Compton scattering of relativistic
electrons off soft photons produced in or in the vicinity of the jet.  However,
the exact role played in the scattering process by photons created in the jet
(synchrotron photons) or outside (broad emission line region or accretion disk
photons) is not known, nor is the cause of the huge, dramatic $\gamma$-ray flares
observed. Depending on which physical parameters are varying (electron
density, magnetic field density, bulk Lorentz factor), variations of different
amplitude are expected in the synchrotron and inverse Compton components (see
Hartman et al. 1996;  Wehrle et al. 1998; Ghisellini \& Madau 1996; Ghisellini \&
Maraschi 1996). The most effective way to discriminate among the different
scenarios consists in observing these sources simultaneously at several
wavelengths spanning a broad range.

PKS~0537--441 ($z = 0.896$) is one of the most luminous and variable blazars at
all frequencies, and has been the target of monitoring from radio to X-rays at
many epochs (Cruz-Gonzalez \& Huchra 1984; Maraschi et al. 1985; Tanzi et al.  
1986; O'Brien et al. 1988; Bersanelli et al. 1992; Edelson et
al.
1992; Falomo et al. 1993a; Treves et al. 1993; Falomo et al. 1994;
Romero et al. 1994; Heidt \& Wagner 1996; Tingay et al. 1996;
Sefako et al. 2000; Tornikoski et al. 2001).  The source was detected by
EGRET for the first time in 1991 (Michelson et al. 1992; Thompson et al. 1993),
and then re-observed at many successive epochs. It is bright and variable in
$\gamma$-rays, and its luminosity in the highest state is comparable to the
average luminosity of the strongest and best studied EGRET blazar, 3C~279.  
Multiwavelength modeling, based on non-simultaneous $\gamma$-ray and lower
frequencies data, has been proposed by Maraschi et al. (1994a)
within the synchrotron self-Compton scheme.

On the basis of its radio variability characteristics (Romero et al.
1995) and an off-centered surrounding nebulosity (Stickel et al. 1988),
it was proposed that PKS~0537--441 is microlensed by stars in a foreground galaxy.  
Search for extended optical emission around PKS~0537--441 has a long and
controversial history (e.g., Falomo et al. 1992; Lewis \& Ibata 2000;
Scarpa et al. 2000). Clarifying the nature of this emission is of importance both
for the study of the properties of the galaxy and environment hosting this very
active nucleus and for the alleged possibility of microlensing effects.

In this paper we present multiwavelength observations of PKS~0537--441 at various
epochs during the EGRET lifetime, and particularly focus on the $\gamma$-ray flare
of 1995 (Sect. 2.1). We also report on 1998 BeppoSAX observations in the 0.2-50 keV
band (Sect. 2.2) and consider archival HST spectra (Sect. 2.3). In Sect. 2.4.2 we
compare
and discuss the results of all imaging studies of the field. In Sect. 3 we discuss
the overall energy distribution and its implications for the nuclear emission
mechanisms.

% --------------     Table 1: JOURNAL OF OBSERVATIONS     -----------------

\begin{table*}[t!] 
\caption[]{Multiwavelength observations of PKS~0537--441}
\begin{center}
\begin{tabular}{lccccc}
\noalign{\smallskip}       
\hline
\noalign{\smallskip}       
Date    & Instrument &  $\alpha_\nu^a$  & $F_\nu^b$   & $\nu^c$    & Ref.$^d$ \\
\noalign{\smallskip}
\hline
\noalign{\smallskip}         
1991 Feb 11      & ESO 1.5m+CCD            & 1.35$\pm$0.05 & 2.19$\pm$0.04 mJy     & 
$5.45 \times 10^{14}$ & 1 \\
1991 Apr 16      & ROSAT+PSPC         & 1.1$\pm$0.4   & 0.79$\pm$0.05 $\mu$Jy & 
$2.41 \times 10^{17}$  & 2 \\
1991 Jul 26-Aug 08 & CGRO+EGRET            & 1.50$\pm$0.32 & 39$\pm$8 pJy         &
$9.66 \times 10^{22}$ & 3,4 \\
1992 May 14-Jun 04 & CGRO+EGRET            & ...           & $< 48^e$ pJy          &
$9.66 \times 10^{22}$  & 5 \\ 
1992 May 21.41     & IUE+LWP         & ...           & 0.66$\pm$0.05$^f$ mJy & 
$1.15 \times 10^{15}$ & 4 \\
1992 May 21.55     & IUE+LWP         & ...           & 0.78$\pm$0.03$^f$ mJy &
$1.15 \times 10^{15}$ & 4 \\
1993 Jul 12        & HST+FOS+G130H     & ...           & 0.09$\pm$0.01$^f$ mJy & 
$2.14 \times 10^{15}$ & 4 \\
1993 Sep 16        & HST+FOS+G270H   & $1.92 \pm 0.09$ & 0.371$\pm$0.007$^f$ mJy 
& $1.15 \times 10^{15}$ & 4 \\
1995 Jan 10-24     & CGRO+EGRET            & 0.96$\pm$0.18 & 137$\pm$21 pJy        & 
$9.66 \times 10^{22}$ & 6 \\
1995 Jan 30.37   & IUE+SWP        & 1.2$\pm$0.1$^g$ & 0.90$\pm$0.02$^f$ mJy & 
$1.67 \times 10^{15}$ & 4 \\ 
1995 Jan 31.44   & IUE+LWP        & 1.2$\pm$0.1$^g$ & 1.45$\pm$0.08$^f$ mJy & 
$1.15 \times 10^{15}$ & 4 \\
1995 Feb 01.17   & ESO SEST          & ...           & 5.16$\pm$0.21 Jy      & 
$90 \times 10^9$ & 4 \\  
1995 Feb 03      & ESO 1.5m+CCD      & 1.29$\pm$0.04 & 3.03$\pm$0.15 mJy     
& $5.45 \times 10^{14}$ & 7 \\
1995 Feb 05      & ESO 1.5m+CCD       & 1.27$\pm$0.05 & 3.64$\pm$0.20 mJy     
& $5.45 \times 10^{14}$ & 7 \\
1995 Feb 07      & ESO 1.5m+CCD       & 1.09$\pm$0.04 & 5.26$\pm$0.25 mJy     
& $5.45 \times 10^{14}$ & 7 \\
1995 Feb 27.14   & ESO SEST                & ...           & 2.88$\pm$0.23 Jy      & 
$230 \times 10^9$ & 4 \\  
1998 Nov 28-30   & BeppoSAX      & 0.80$\pm$0.13 & 0.46$\pm$0.09 $\mu$Jy & 
$2.41 \times 10^{17}$ & 4 \\
\noalign{\smallskip}
\hline
\noalign{\smallskip}
\multicolumn{6}{l}{$^a$ Spectral index ($F_\nu \propto \nu^{-\alpha_\nu}$).
Uncertainties, both for $\alpha$ and fluxes, are at 90\% confidence level}\\
\multicolumn{6}{l}{ ~~ for the X- and $\gamma$-ray measurements, and at 68\% for UV, 
optical and millimetric. No spectral fit has been}\\
\multicolumn{6}{l}{ ~~ tried when the data signal-to-noise ratio was too low or the
spectral range too limited.}\\
\multicolumn{6}{l}{$^b$ Flux density. EGRET flux conversion follows Thompson et al. 
(1996).
Optical-to-X-ray data are corrected}\\
\multicolumn{6}{l}{ ~~ for Galactic extinction.}\\
\multicolumn{6}{l}{$^c$ Frequency to which the flux density 
refers, in Hz.}\\
\multicolumn{6}{l}{$^d$ References. 
{\bf 1:} Falomo et al. 1994;
{\bf 2:} Treves et al. 1993;
{\bf 3:} Thompson et al. 1993;}\\
\multicolumn{6}{l}{ ~~ {\bf 4:} This paper; {\bf 5:} Hartman et al. 1999;
{\bf 6:} Mukherjee et al. 1997;
{\bf 7:} Scarpa \& Falomo 1997.}\\
\multicolumn{6}{l}{$^e$ 2-$\sigma$ upper limit.}\\
\multicolumn{6}{l}{$^f$ The uncertainty is only statistical. For IUE data, this was
evaluated
following Falomo et al. (1993b).}\\
\multicolumn{6}{l}{$^g$ From a spectral fit over the band 1700-5500 \AA.}\\
\end{tabular}
\end{center}
\end{table*}                              

\section{Observations, data analysis and results in individual bands}
\label{}

\subsection{$\gamma$-rays}

PKS~0537--441 was repeatedly observed in the band 0.03-10 GeV by EGRET aboard
CGRO: a summary of the observations, with flux levels or upper limits, is reported
in the 3rd EGRET catalog (Hartman et al. 1999).  We have re-analyzed the
July-August 1991 spectrum with up-to-date software and obtain a marginally steeper
slope (see Table~1) than found by Thompson et al. (1993). During January 1995 the
source was detected with almost 10-$\sigma$ significance, with an average flux of
$90 \times 10^{-8}$ photons s$^{-1}$ cm$^{-2}$, its brightest recorded
$\gamma$-ray state (see Hartman et al. 1999 for details of the data reduction and
analysis).  The high signal-to-noise ratio of the detection allowed a rather
detailed spectral analysis (see Mukherjee et al. 1997 and Table 1) and a study of
the variability on a day time scale. In Fig. 1 we report the EGRET light curve of
the blazar during the January 1995 pointing, with a temporal binning of 2 days
(see also Hartman 1996).  Toward the end of the monitoring, the flux increases in
$\sim$2 days by a factor of at least $\sim$3 with respect to the average level
during the first 10 days of the observation.  The target was detected by EGRET
also three months later, in a state $\sim$4 times lower than shown here (Hartman
et al.  1999).

In Table 1 and Fig. 2 we report the EGRET data taken quasi-simultaneously (within
$\sim$6 months)  with data at lower frequencies. The January 1995 spectrum is
significantly flatter than measured in 1991, consistent with a common
characteristics of $\gamma$-ray blazars, which tend to have flatter spectra during
flare states (Sreekumar et al. 1996; Sreekumar et al. 2001).

% FIGURE 1: EGRET LIGHT CURVE

\begin{figure}
\psfig{file=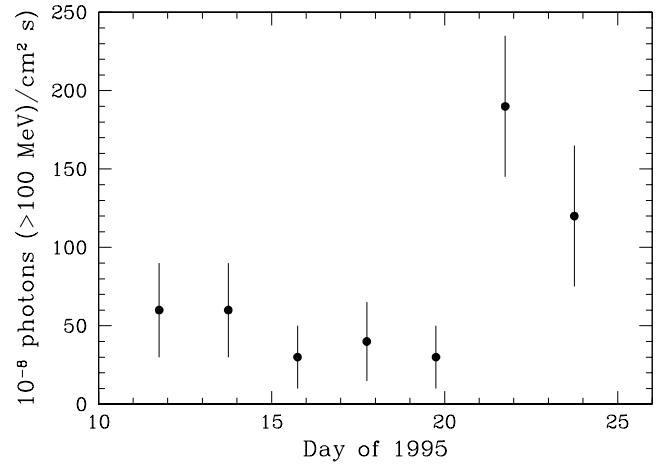,width=9.5cm}
\caption{EGRET light curve of PKS~0537--441 in January 1995}
\end{figure}

\subsection{X-rays}

X-ray observations of PKS~0537--441 prior to 1991 are summarized by Treves et al.
(1993).  The results of the ROSAT 1991 observation presented by those authors are
reported in Table 1 and Fig. 2.

We observed PKS~0537--441 with BeppoSAX in 1998 November 28.3142-30.5291 (UT)  as
part of a program focussed on BL Lacs with flat X-ray spectra, supposedly
dominated by the inverse Compton component. The total on-source integration times
were 23672, 47396, and 36743 seconds for the LECS (0.1-4 keV), MECS (1.6-10 keV)
and PDS (13-300 keV) instruments, respectively (see Scarsi 1993 and Boella et al.
1997 for overviews of the BeppoSAX satellite).

% FIGURE 2: MW SED

\begin{figure*}
\psfig{file=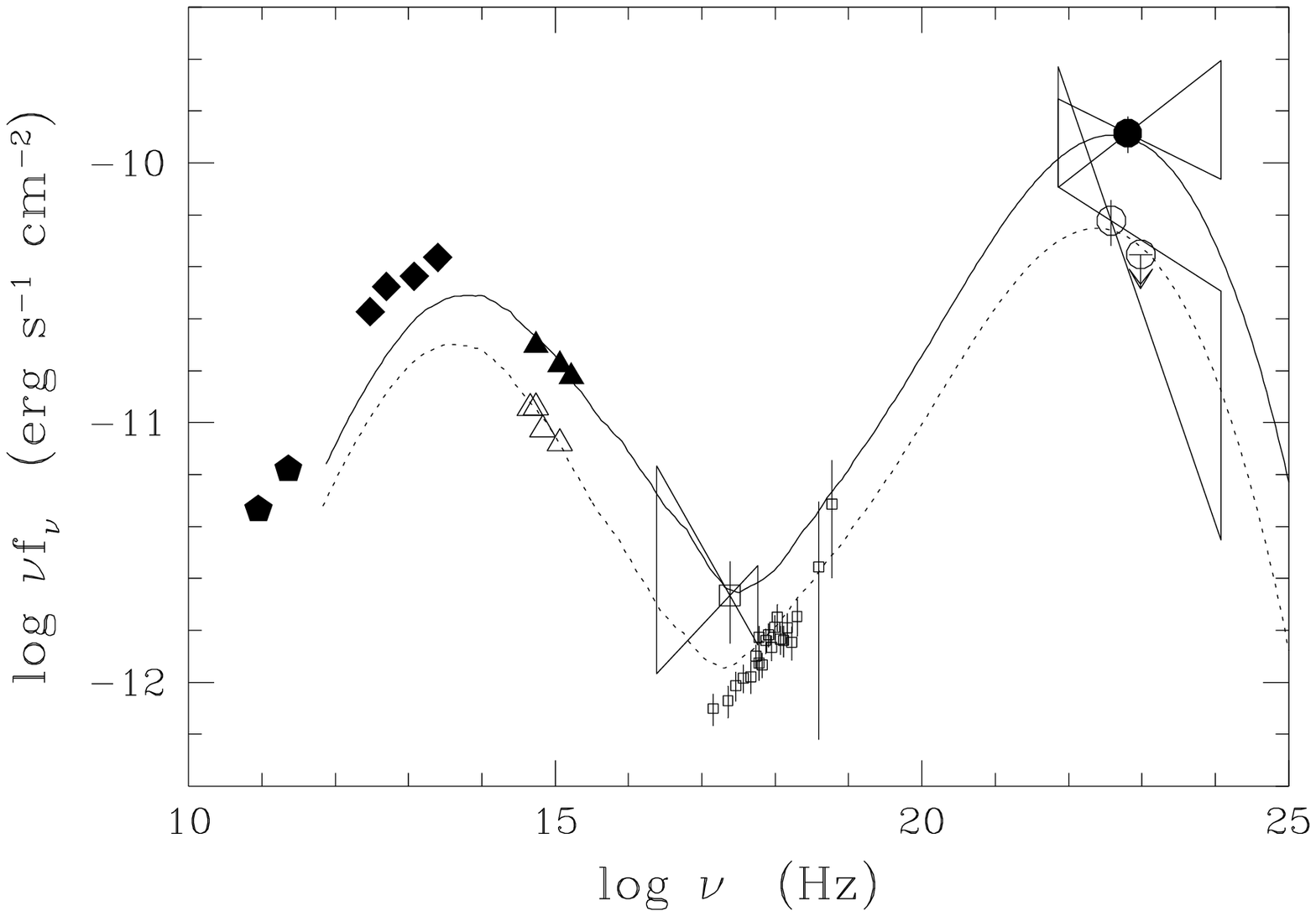,width=16cm}

\caption{Radio-to-$\gamma$-ray spectral energy distribution of PKS~0537--441 at
different states (see Table 1 for data references).  The filled pentagons,
triangles and circle represent millimetric, optical-to-UV and $\gamma$-ray data in
Jan-Feb 1995, respectively; diamonds are IRAS 1984 data from Impey \& Neugebauer
(1988). The open triangles stand for optical February 1991 and UV 1992 data, open
squares represent ROSAT April 1991 (large symbol) and BeppoSAX 1998 (small symbol)
data, and open circles represent the EGRET Jul-Aug 1991 detection and 1992 upper
limit. Data at optical-to-X-ray frequencies are corrected for Galactic extinction
(see text). Superimposed on the multiwavelength data are the model curves for the
low 1991-1992 state (dotted), and for the high 1995 state (solid).  The model
parameters for the low state are $\gamma_{\rm min} = 1$, $\gamma_{\rm b} = 400$,
$\gamma_{\rm max} = 2.5 \times 10^4$, $n_1 = 1.6$, $n_2 = 4$, $B = 3.5$ Gauss, $K = 6
\times 10^3$ cm$^{-3}$, $R = 2.7 \times 10^{16}$ cm, $\delta \simeq \Gamma = 10$
(see Ballo et al. 2002 for the parameter notation).  In high state, the above
parameters remain unchanged, except that $\Gamma = 11$ and $n_2 = 3.8$. The
external Compton seed photon source (disk photons reprocessed in the broad line
region) is assumed to have a density $U_{\rm ext} = 6 \times 10^{-3}$ erg
cm$^{-3}$; the disk temperature is assumed to be $10^4$ K}

\end{figure*}

The spectra taken by the LECS and MECS instruments have been extracted from the
linearized event files using radii of 8$^{\prime\prime}$ and 4$^{\prime\prime}$,
respectively, and corrected for background contamination with library files
available at the BeppoSAX Science Data Center. The net count rates, averaged over
the whole pointing, are (1.93$\pm$0.11)$\times 10^{-2}$ cts s$^{-1}$ in the LECS
and (2.37$\pm$0.08)$\times 10^{-2}$ cts s$^{-1}$ in the MECS.  The spectra
measured by the PDS instrument have been accumulated from on source exposures of
the collimator units, and corrected using background spectra obtained during the
off-source exposures. A correction for the energy and temperature dependence of
the pulse rise time has been also applied (Frontera et al. 1997). The average PDS
spectrum of the source exhibits significant signal up to $\sim$30 keV and appears
featureless; its net count rate is (3.9$\pm$1.6)$\times 10^{-2}$ cts s$^{-1}$.  
No emission variability larger than 15\% has been detected within the pointing.

After rebinning the LECS, MECS and PDS spectra in intervals where the signal
exceeds a 3-$\sigma$ significance, we fitted them to a single absorbed power-law
using the XSPEC routines and response files available at the Science Data Center.  
The LECS and PDS spectra normalizations relative to the MECS have been treated as
a free (best fit value 0.86$\pm$0.09, within the expected range) and fixed (0.85)  
parameters, respectively, and the neutral hydrogen column density has been fixed
to its Galactic value ($N_{\rm HI} = 2.91 \times 10^{20}$ cm$^{-2}$, Murphy et al.
1996).  The fit is satisfactory ($\chi^2_\nu = 0.6$).  The best-fit power-law
parameters are reported in Table 1 and the deconvolved de-absorbed X-ray fluxes in
the BeppoSAX energy band are reported in Fig. 2. We also searched for a possible
iron emission line corresponding to the K$\alpha$ transition at the expected
redshifted energy on the MECS spectrum, but we do not detect any line with
intensity larger than $1.4 \times 10^{-14}$ erg s$^{-1}$ cm$^{-2}$ (3$\sigma$
upper limit).

The spectrum and emission state detected by BeppoSAX are consistent with those
measured at earlier epochs by EXOSAT and Einstein (Treves et al.  1993), while the
BeppoSAX flux at 1 keV is almost a factor of $\sim$2 lower than that detected by
ROSAT in 1991.

\subsection{Ultraviolet}

IUE performed low dispersion spectroscopy of PKS~0537--441 in the 2000-3000 \AA\
(LWR or LWP cameras) and 1200-1950 \AA\ (SWP) wavelength ranges at various epochs
between 1980 and 1995, and in two occasions, May 1992 and January 1995,
quasi-simultaneously with EGRET pointings (Table~1). The corresponding IUE spectra
were retrieved from the
INES\footnote{http://ines.vilspa.esa.es/ines/docs/contents.html} and
NEWSIPS\footnote{http://archive.stsci.edu/iue/index.html} (Nichols \& Linsky 1996;
Garhart et al. 1997) archives and, for the 1995 data only, extracted from the
bi-dimensional spectral images with the GEX routine (Urry \& Reichert 1988).

The INES and NEWSIPS spectra of May 1992 agree in flux to within $\sim$10\%, and
therefore can be considered consistent, given the low level of the source and the
possible differences in background estimate between the two extraction methods. In
Table 1 and Fig. 2 we have reported the results obtained with the NEWSIPS
extraction.  As a consistency check, we note that these UV fluxes are compatible
with the power-law fitting the optical data taken more than one year earlier (see
Table~1 and Fig. 2).

For the spectra of January 1995, the INES and NEWSIPS methods return compatible
results for the IUE SWP range (the flux at 1800 \AA, averaged between the two
extractions and corrected for Galactic extinction, is 1.21$\pm$0.03 mJy),
while in the LWP range the NEWSIPS extracted flux is up to a factor of 2 higher
than found with INES.  The GEX extracted flux is $\sim$25\% lower than that given
by INES and NEWSIPS in the SWP band, and is consistent with the INES result in the
LWP band.  Since we have verified that the differences between pairs of IUE
spectra of bright sources (e.g., the BL Lac object PKS~2155--304) extracted with
different methods do not exceed 1-2\% (see also Urry et al. 1993), we ascribe the
discrepancies we find for PKS~0537--441 to the difficulty of correctly evaluating
the background at UV flux levels comparable with the IUE sensitivity limit and of
properly removing the solar scattered light, which critically contaminates the IUE
LWP camera acquisitions after 1992 (Caplinger 1995, and references therein).

Unlike the NEWSIPS and INES spectra, the fluxes resulting from the GEX spectra of
PKS~0537--441 are consistent with the power-law which fits the quasi-simultaneous
(within few days) optical data (see also below). Therefore, we have decided to
adopt the GEX extraction for the 1995 IUE spectra. These have been calibrated
according to Bohlin et al. (1990, SWP) and Cassatella et al. (1992, LWP), and have
been binned in 70-100 \AA-wide wavelength intervals, after removing cosmic rays
and by avoiding regions affected by spurious sunlight ($\lambda \simgt 2800$ \AA)
and by large noise (1900 \AA $\simlt \lambda \simlt 2300$ \AA).  The binned
spectra are reported in Fig. 3.

% FIGURE 3: UV SPECTRAL CONTINUA

\begin{figure}
\psfig{file=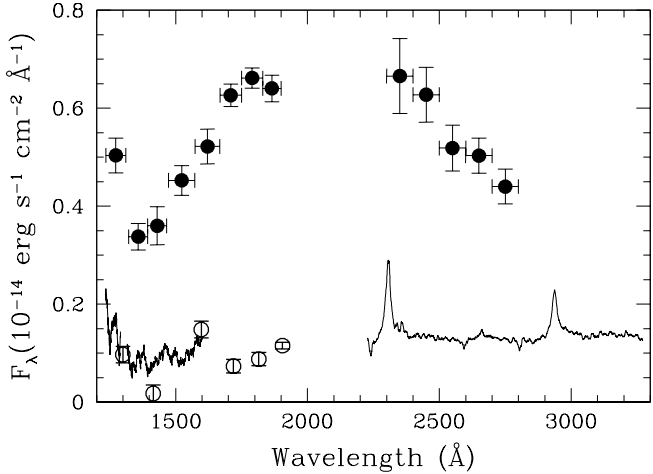,width=9.5cm}

\caption{IUE (circles) and HST FOS (solid curves) spectra of PKS~0537--441 not
corrected for Galactic extinction.  The IUE data refer to a bright (January 1995,
filled circles) and a faint (April 1982, open circles) state. The FOS G130H and
G270H spectra, smoothed over wavelength intervals of 12.5 and 5 \AA,
respectively, sample a low state (July and September 1993, respectively),
comparable to that observed by IUE in 1982.  Note that the strongest emission
lines, Ly$\alpha$ and C~IV, well visible in the HST spectrum, occur out of the
usable wavelength region of the IUE data}

\end{figure}

No emission lines superimposed on the UV continuum are detected in any of the IUE
spectra.  In particular, the Ly$\alpha$ and C IV lines, well detected by HST (see
below), fall at the edges of the LWP range, where the sensitivity is poor.
Moreover, the high state detected by IUE in 1995 (a factor of $\sim$4 higher than
observed by HST), may have implied smaller line equivalent widths.  A suggestion
of C~IV emission may be present in the average of the LWP spectra of PKS~0537--441
taken before 1988 (Courvoisier \& Paltani 1992).

A power-law with slope $\alpha_\nu \sim 1.2$ ($f_\nu \propto \nu^{-\alpha_\nu}$)
fits marginally well (reduced $\chi^2 \simeq 2$) the 1995 optical and UV data at
wavelengths longer than $\sim$1700 \AA, corrected for Galactic dust extinction
($E_{\rm B-V} = 0.037$ mag, Schlegel et al. 1998) with the law by
Cardelli et al.  (1989).  The spectral index is in agreement with that
determined for the optical spectra only (Scarpa \& Falomo 1997).  The LWP spectrum
appears steeper than this, which may account for the not completely satisfactory
value of the $\chi^2$.  The discrepancy may be due to the limited quality of the
IUE LWP data, and to the non-strict simultaneity.

The UV points at wavelengths shorter than 1700 \AA\ clearly deviate from the
optical-to-UV power-law behavior and suggest a flux deficit by up to 50\% at
$\sim$1350 \AA\ (Fig. 3).  This feature is seen independent of the method adopted
for spectral extraction. However, we note that the background has an irregular
shape along the dispersion direction, and possible instrumental effects cannot be
excluded. The only other SWP spectrum available for PKS~0537--441 (25 April 1982,
Fig. 3) obtained at a very faint state, has insufficient signal-to-noise ratio to
confirm or disprove the reality of the feature (see also Lanzetta et al. 1995).
Moreover, a similar dip is not seen in the HST FOS G130H spectrum (see below),
which has a level comparable to that of the 1982 IUE spectrum (Fig. 3), but a much
better signal-to-noise ratio. Therefore, we do not consider this feature in the
analysis of the multiwavelength energy distribution (Sect. 3).

% FIGURE 4:  HST FOS G270 SPECTRUM

\begin{figure}
\psfig{file=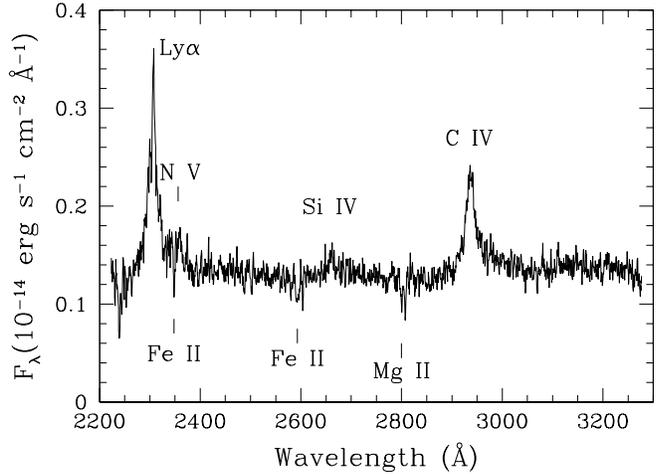,width=9.5cm}
\caption{FOS spectrum (G270H) of the blazar PKS~0537--441.
Strong nuclear
emission lines  ($z = 0.896$) and faint Galactic absorption features are clearly
detected}
\end{figure}

PKS~0537--441 was observed by the HST Faint Object Spectrograph (FOS)  in July and
September 1993 (pre-COSTAR) with the G130H and G270H gratings, respectively
(Table~1).  We retrieved both spectra from the HST
archive\footnote{http://archive.eso.org/wdb/wdb/hst/science/form} (see also Bechtold et
al. 2002). The G130H spectrum, which is rather noisy, presents a strong, sharp
emission feature at $\sim$1300 \AA\ that cannot be identified with any known
emission (if at rest frame $\lambda \sim 690$ \AA).  Because of this and the
unusual shape we believe it is an instrumental effect.  On the other hand, the
G270H spectrum has a good signal-to-noise ratio, and exhibits broad, intense
emission lines and Galactic interstellar absorptions (see Fig. 4).

We derived the central wavelength of each emission line by fitting the core of the
features with Gaussians.  Since the Ly$\alpha$ profile is slightly asymmetric due
to a possible narrow absorption on the blue side and close to the center of the
line, we derived the centroid of this line by excluding the inner portion of the
core. On the basis of their measured central wavelengths (see Table~2), the
emission lines are consistently identified with Ly$\alpha$, Si IV, and C IV at $z
= 0.896 \pm 0.001$. This is slightly larger than, but still consistent with
historical redshift determination ($z = 0.894$, Peterson et al. 1976), based on
Mg~II$\lambda$2798 emission, but only marginally consistent with $z = 0.892 \pm
0.001$ given by Lewis and Ibata (2000), based on the Balmer line series and O~III.
The intensities and equivalent widths of the UV emission lines are reported in
Table~2. After excluding the spectral regions affected by the lines, the UV
continuum is well fitted by a power-law of index $\alpha \simeq 1.9$.

%  --------------------------   TABLE 2  ------------------------------

\begin{table}
\caption[]{UV Emission Lines of PKS~0537--441}
\begin{center}
\begin{tabular}{lccc}
\hline
\hline
Ion & $\lambda^a$ & EW$^b$ & I$^c$ \\
\hline 
Ly$\alpha$ $\lambda$1216 & $2305.9 \pm 0.5$ & $30 \pm 5$ & $5.5 \pm 0.8$  \\
Si IV $\lambda$1403      & $2660.0 \pm 0.5$ & $1.5 \pm 0.2$ & $0.25 \pm 0.04$ \\
C IV $\lambda$1549       & $2937.0 \pm 0.5$ & $22 \pm 3$ & $3.6 \pm 0.5$ \\
\hline
\multicolumn{4}{l}{$^a$ Observed line wavelength in \AA.}\\
\multicolumn{4}{l}{$^b$ Observed equivalent width in \AA.}\\
\multicolumn{4}{l}{$^c$ Line intensity in $10^{-14}$ erg s$^{-1}$
cm$^{-2}$, corrected for}\\
\multicolumn{4}{l}{ ~ Galactic extinction.}\\
\end{tabular}
\end{center}
\end{table}

We computed the continuum fluxes from the HST and IUE spectra by averaging the
signal in 200-\AA-wide bands in spectral regions free from lines, spurious
features and noise, and corrected for Galactic extinction (see Table~1 and Fig.
2). The UV state of 1995 was about a factor 2 higher than in 1992 and a factor 4
higher than in 1993.  The optical-to-UV spectrum in 1995 was flatter than the HST
UV spectrum in 1993, consistent with the expected correlation of spectral
flattening with brightening in synchrotron sources.

\subsection{Optical}

\subsubsection{Spectroscopy}

Optical CCD spectrophotometry was obtained in February 1991 (Falomo et al. 1994)  and
on 3,5 and 7 February 1995 at the 1.5m telescope of the European Southern Observatory
(see Scarpa \& Falomo 1997 for details of data acquisition, reduction and analysis,
and Table~1 and Fig. 2 for the results).  A 70\% increase is apparent between 1991 and
1995 in the optical flux.  We detected a monotonic increase of about 70\% during the 3
nights of the 1995 monitoring, accompanied by spectral hardening.  The Mg
II$\lambda$2798 emission line is observed in all spectra, with non-significantly
variable intensity, $I_\lambda \simeq 6 \times 10^{-15}$ erg s$^{-1}$ cm$^{-2}$ (not
corrected for Galactic reddening, Scarpa \& Falomo 1997; Treves et al. 1993.  Note
that the line intensity in the latter reference is misreported by a factor of 10).

\subsubsection{Imaging}

Stickel et al. (1988) reported the detection of extended optical emission
surrounding PKS~0537--441, the characteristics of which led them to interpret it
as a foreground disk galaxy along the line of sight to the blazar. However, this
finding was not confirmed either by higher resolution optical images (Falomo et
al. 1992) and near-IR imaging (Kotilainen et al. 1998), consistent with
the high redshift of the source and the relatively bright nucleus. This source was
also imaged by HST using WFPC2 and F702W filter (Urry et al. 2000; Scarpa et al.
2000) and found to be unresolved. However, based on the analysis of the same HST
images, Lewis and Ibata (2000) claim the finding of faint non-axisymmetric
extended emission.  This result was obtained from the subtraction of a scaled PSF
constructed with the Tiny Tim software (Krist 1995). While Tiny Tim offers a
reliable description of the central part of the HST-WFPC2 PSF it does not account
for the emission due to the large angle scattered light (particularly for the
Planetary Camera) that produces extra emission at radii larger than $\sim$2
arcsec, as illustrated in Fig. 5 (see also Scarpa et al. 2000).  We therefore
believe that the extra light reported by Lewis and Ibata (2000) is fully accounted
for by the contribution of the scattered light.  Moreover, we note that no
signature of stellar features has been reported in the optical spectra of
PKS~0537--441 to support the presence of a foreground galaxy (Stickel et al. 1988;
Lewis \& Ibata 2000).  Therefore, the microlensing hypothesis does not appear
tenable (see also Heidt et al. 2002).

\subsection{Millimeter}

PKS~0537--441 was monitored during the year 1995 with the 15-metre
Swedish-ESO Submillimetre Telescope (SEST) at La Silla, Chile (Tornikoski et al.
2002, in preparation). Observations were made at 90 GHz using a dual polarization
Schottky receiver, and at 230 GHz using a single channel bolometer. More details
on the observations and data reduction can be found in Tornikoski et al. (1996).  
The sampling is only marginally simultaneous with EGRET.  The fluxes obtained
closest in time with the EGRET observation are reported in Table~1, along with
errors, which consist of the standard deviation among the individual daily
measurements and the estimated effect of the calibration and pointing
uncertainties.

These data suggest a very moderate brightening in the radio flux curve at 90 GHz
just about the time of the multiwavelength campaign, but not comparable to the
1994 outburst (Tornikoski et al. 1996).  In the 230 GHz data no changes can be
seen from December 1994 to the end of February 1995, but the sampling is too
sparse to completely exclude the possibility of a minor burst similar to that seen
in the 90 GHz curve (see Tornikoski et al. 1996).

% FIGURE 5:  HST WFPC2 IMAGES AND PROFILES

\begin{figure*}
%\psfig{file=H3499F5.eps,width=16cm}

%\vspace{0.5cm}
\vspace{5cm}

\caption{The upper panels show, from left to right, the HST Planetary Camera images in
F702W filter of PKS~0537--441, the unresolved blazar H~1722+119, and a star (the angular
size of each field is $9^{\prime\prime}.2 \times 9^{\prime\prime}.2$). The lower panels
report, from left to right, the PSF model fit (solid curve), including the contribution
of scattered light, to the observed profiles of surface brightness (in mag arcsec$^{-2}$)
vs radius (in arcsec) for PKS~0537--441, H~1722+119, and many stars (dots).  In the third
lower panel we also report, for comparison, the Tiny Tim PSF model (dashed curve), which
is clearly unable to describe the profile of point-like sources at radii larger than
$\sim$2 arcsec (see also Scarpa et al. 2000)}

\end{figure*}

\section{Discussion}
\label{}

In Fig. 2 we have reported the broad-band spectral energy distributions of
PKS~0537--441 constructed with quasi-simultaneous data from the millimetric to the
$\gamma$-ray frequencies (see Table 1).  In addition, we have reported the
unpublished BeppoSAX spectrum.  The multiwavelength state of PKS~0537--441 in
January-February 1995 was one of the brightest recorded for this source during the
lifetime of EGRET (cf. Tornikoski et al.  1996; Falomo et al. 1994; Hartman et al.
1999). The comparison of $\gamma$-ray detection and optical flux of 1991 and the
EGRET upper limit and UV flux of 1992 is suggestive of little or no
multiwavelength variability between the two epochs.  The increase of the
optical-to-UV flux from the 1991-1992 to the 1995 level corresponds to a variation
of similar amplitude (a factor of $\sim$2) in the $\gamma$-rays, or only slightly
larger. The flat BeppoSAX spectrum suggests that a single emission component
dominates in the energy band 0.1-30 keV.

PKS~0537--441 exhibits the typical double-humped multiwavelength spectral shape of
``Low-frequency-peaked" blazars (Padovani \& Giommi 1995; Sambruna et al. 1996;
Fossati et al. 1998): the first component peaks at wavelengths longer
than the optical (likely in the far-infrared, as suggested by the IRAS data, taken
at a much earlier epoch, Impey \& Neugebauer 1988) and is due to synchrotron
radiation.  The second component, which peaks around $\sim 10^{22}-10^{24}$ Hz
(Fig. 2) and is a factor $\sim$6 more powerful than the synchrotron maximum, is
probably produced via inverse Compton scattering between relativistic electrons
and synchrotron photons (Maraschi et al. 1992; Maraschi et al. 1994b; Bloom \&
Marscher 1993) or external photons (broad emission line region or accretion disk,
Dermer \& Schlickeiser 1993; Sikora et al. 1994).

The overall energy distribution of PKS~0537--441, the inverse-Compton dominance,
and the presence of optical and UV emission lines suggest that external Compton
upscattering may be significant with respect to the synchrotron self-Compton
process in producing the high energy emission (Ghisellini et al. 1998; Ghisellini
2001). To model the broad-band energy distribution of PKS~0537--441 we have
assumed that the emission is produced in a region filled by relativistic particles
which radiate at low energies via synchrotron, and upscatter both synchrotron
photons and accretion disk photons reprocessed in the broad emission line region.  
We have not included the direct contribution of the accretion disk to the
population of seed photons to be upscattered by the jet electrons, because those
would be highly redshifted in the frame of the blob, which is emitting at a
distance of $\sim$ 0.3 pc (see caption to Fig. 2) from the jet base (Sikora et al.
1994). The size of the emitting region has been constrained with the variability
time scale in $\gamma$-rays ($\sim$2 days); the electron energy distribution is
modeled with a double power-law (see Tavecchio et al. 2000 and Ballo et al. 2002
for more details of the model). The low state model has been constrained with the
BeppoSAX spectrum and with the 1991-1992 optical and $\gamma$-ray data, while the
high state model reproduces the 1995 multiwavelength data.  Note that the ROSAT
spectral measurement of April 1991, which is affected by a large uncertainty, is
consistent both with the low and high state model curves. The IRAS data are 
marginally consistent with the high state model.

The difference between the two model curves (reported in Fig. 2; see caption for
the model parameters) is solely determined by a change in the Lorentz factor of
the relativistic plasma bulk motion (increasing by 10\% from the low to the high
state) and in the index of the upper branch of the electron energy distribution
($n_2$ in the notation of Ballo et al. 2002, slightly flatter in the brighter
state, see caption to Fig. 2).  Under our assumption that the high energy
component results from the contribution of both synchrotron-self Compton and
external Compton mechanisms, the variation of the bulk Lorentz factor would
produce a correlation between the variability amplitude of the synchrotron and
inverse Compton components which is intermediate between linear and half-cubic
(Ghisellini \& Maraschi 1996). This is consistent with the observations, although
the difficulty of exactly locating the peaks of the emission components, the
non-strict simultaneity of the optical-to-UV spectra and the rapid $\gamma$-ray
flare seen at the end of the EGRET viewing period make the fit results only
indicative.

From our measured UV line intensities reported in Table~2 and from dereddened MgII
emission line intensity (Sect. 2.4.1), assuming $H_0$ = 65 km s$^{-1}$ Mpc$^{-1}$,
$\Omega_{\rm m} = 0.3$, $\Omega_\Lambda = 0.7$, we derive a total line luminosity $L_{\rm
BLR} \sim 4.7 \times 10^{44}$ erg s$^{-1}$ (the other observed optical emission lines are
sufficiently weak that their contribution to the line luminosity is not significant, see
Lewis \& Ibata 2000). Using this and the external photon density assumed in our model,
$U_{\rm ext} = 6 \times 10^{-3}$ erg cm$^{-3}$, we can evaluate the size of the broad
line region, $R_{\rm BLR} \simeq 7.9 \times 10^{17}$ cm.  As an independent check, we
have considered the empirical relationship determined by Kaspi et al. (2000) between the
size of the broad line region and the luminosity of the thermal continuum at 5100 \AA\ in
quasars.  Assuming a covering factor of the broad line clouds of 10\%, we can estimate
the disk luminosity to be $L_{\rm disk} \sim 4.7 \times 10^{45}$ erg s$^{-1}$.  This
value is consistent with the approximate upper limit which can be derived from the broad
band spectrum, $\sim 2 \times 10^{46}$ erg s$^{-1}$.  After accounting for a factor of
$\sim$3 difference between a bolometric and a monochromatic disk output, we find that our
disk luminosity estimate implies, according to Kaspi et al.'s formula, $R_{\rm BLR}
\simeq 5.8 \times 10^{17}$ cm, well consistent, given the uncertainties, with the size we
have determined based on the observed $L_{\rm BLR}$ and on the assumed photon density. We
finally note that our observed $L_{\rm BLR}$ is about a factor 4 lower than that
estimated for this source by Celotti et al. (1997), based on an old measurement of the
MgII emission only.

The validity and origin of the spectral dip seen in the IUE spectrum of January
1995 at wavelengths shorter than 1700 \AA\ remain to be established. Assuming it
to be real, it could be qualitatively described by an absorption edge or broad
trough, and may be identified with neutral hydrogen ionization discontinuity
(Lyman limit)  at the redshift of the source or with Ly$\alpha$ forest blanketing
up to a maximum redshift of $\sim$0.4.  Continuum decrements shortward of
Ly$\alpha$ emission, Lyman limit systems and damped Ly$\alpha$ absorption features
have been reported in few blazars at redshifts $0.5 \simlt z \simlt 1.5$ (e.g.,
PKS~0637--752, Cristiani et al. 1993; PKS~0735+178, PKS~2223-052 Courvoisier \&
Paltani 1992, Lanzetta et al. 1995).  Absorption features have recently been
reported also in X-ray blazar spectra (Tavecchio et al. 2000). The host galaxy or
the halos of the galaxies located in the vicinity of PKS~0537--441 and at similar
redshift (projected distances of $\sim$30 kpc, Heidt et al. 2002) may be
responsible for absorption in the bluer part of the UV nuclear spectrum.  
Using the photoelectric cross-section computed by Rumph et al. (1994)
to model the interstellar opacity at extreme UV wavelengths, we estimated that in
the case of Lyman continuum absorption the equivalent intrinsic $N_{\rm HI}$ would be
about 3 orders of magnitude lower than the Galactic $N_{\rm HI}$ in the direction of
the blazar, and therefore its effect on the X-ray spectrum would be undetectable.
Alternatively, the FOS G130H spectrum may suggest resemblance with those of Broad
Absorption Line quasars (e.g., Hamann \& Ferland 1999; Arav et al. 2001), so that,
in the lower resolution IUE SWP spectrum, many broad features would appear blended
in a unique edge.

Higher signal-to-noise ratio data (ideally HST STIS spectra simultaneously
covering the wavelength range 1100-2700 \AA) would be necessary to confirm the
presence of the feature and accurately model it. The nuclear emission and
environmental characteristics of the blazar PKS~0537--441 make it a case study for
future $\gamma$-ray missions like AGILE and GLAST and for the NGST, respectively.

\begin{acknowledgements}

We thank W. Wamsteker and the IUE Observatory staff for their support of this
project, C. Morossi for useful comments on the IUE spectral response, C. Imhoff,
K. Levay and P. Padovani for assistance with IUE and HST archives, and P.
Goudfrooij, M. Massarotti, P. Molaro, and M. Stiavelli for helpful suggestions.
This work is partly supported by COFIN 2001/028773, ASI-IR-35. This research has
made use of the NASA/IPAC Extragalactic Database (NED) which is operated by the
Jet Propulsion Laboratory, California Institute of Technology, under contract with
the National Aeronautics and Space Administration.

\end{acknowledgements}

% ---------------------    REFERENCES    ------------------------

\end{document}